\documentclass[journal=apchd5,manuscript=letter, layout=twocolumn]{achemso}

\usepackage[version=3]{mhchem} 

\author{Marco Barbieri}
\affiliation{Dipartimento di Scienze, Universit\`a degli Studi Roma Tre, Via della Vasca Navale 84, 00146 Rome, Italy}
\alsoaffiliation
{Istituto Nazionale di Ottica - CNR, Largo Enrico Fermi 6, 50125 Florence, Italy}
\email{marco.barbieri@uniroma3.it}

\author{Iole Venditti}
\affiliation{Dipartimento di Scienze, Universit\`a degli Studi Roma Tre, Via della Vasca Navale 84, 00146 Rome, Italy}

\author{Chiara Battocchio}
\affiliation{Dipartimento di Scienze, Universit\`a degli Studi Roma Tre, Via della Vasca Navale 84, 00146 Rome, Italy}

\author{Vincenzo Berardi}
\affiliation{Dipartimento Interateneo di Fisica ''M.~Merlin", Politecnico di Bari, Via Orabona 4, 70126 Bari, Italy}

\author{Fabio Bruni}
\affiliation{Dipartimento di Scienze, Universit\`a degli Studi Roma Tre, Via della Vasca Navale 84, 00146 Rome, Italy}

\author{Ilaria Gianani}
\affiliation{Dipartimento di Scienze, Universit\`a degli Studi Roma Tre, Via della Vasca Navale 84, 00146 Rome, Italy}

\title{Observing thermal lensing with quantum light}


\begin{document}


\begin{abstract}
  The introduction of quantum methods in spectroscopy can provide  enhanced performance and technical advantages in the management of noise. We investigate the application of quantum illumination in a pump and probe experiment. Thermal lensing in a suspension of gold nanorods is explored using a classical beam as the pump and the emission from parametric downconversion as the probe. We obtain an insightful description of the behaviour of the suspension under pumping with a method known to provide good noise rejection. Our findings are a further step towards investigating effects of quantum light in complex plasmonic media.
\end{abstract}

\section{}
Manipulation and control of quantum light can give way to new approaches in the investigation of matter and materials~\cite{RevModPhys.88.045008,Mukamel_2020,MarcusRaymer}. Quantum optical techniques are known to offer superior performance in metrology when compared to classical light with the same energy~\cite{BraunsteinCaves,Dowling,GLM,Polino}. Their employ is mostly sought for the investigation of fundamental effects in the absorption-emission process~\cite{Raymer2D,RamanMukamel,PhysRevLett.112.213601,PhysRevLett.129.183601,Li23}. At the same time, it has also been appreciated that using quantum or quantum-inspired methods can provide a technical advantage~\cite{Scarcelli,Yabushita04,Hosten,PhysRevX.4.011031,PhysRevX.4.011049,Kalashnikov,Steinberg,Peter,PhysRevLett.125.080501}, even when a fundamental improvement is put into question~\cite{PhysRevLett.114.210801,PhysRevLett.116.040502}.

More recently, quantum light has  been employed to study plasmonic materials\cite{Tame,PhysRevLett.101.190504,Messin1,Messin2,PhysRevLett.121.173901,PhysRevX.4.011049} leading to sensing applications in which a convenience in coupling quantum light to plasmonic structures is demonstrated~\cite{Dowran:18,Chen:18, Lee:18}. Along with structured materials, suspensions of nanoparticles and nanorods can also present plasmonic effects that are exploited in spectroscopy~\cite{ChemReview,C1AN15313G}. These can also be observed in non-stationary conditions under pumping by means of time-resolved thermal lensing. This is indeed a standard technique to investigate solutions~\cite{Pedrosa:20,AgiotisMeunier}. It is hard to foresee a successful application of quantum enhanced sensing, as these systems are intrinsically lossy due to scattering. In fact, loss is one of the most detrimental factors in quantum sensing~\cite{Rafal}. Yet, practical advantages could still be investigated.

An intriguing suggestion in this sens comes from the domain of quantum imaging. Multimode quantum light, presenting intensity correlations, is useful for identifying the presence of an object despite high levels of noise. The proposal, termed quantum illumination~\cite{Lloyd08}, has been the subject of experimental and theoretical investigations~\cite{PhysRevLett.101.253601,PhysRevLett.110.153603,PhysRevLett.114.110506,gregory,Nair:20}. In this article we explore the application of these concepts in a time-resolved measurement of thermal lensing ina suspension of gold nanorods (AuNR). This is induced by ultrashort pulses and probed by quantum light. The use of quantum correlations provides a useful time trace, with potential for noise rejection, transposing their employ from the spatial~\cite{Gili22} to the time domain. We have tested the technique in standard conditions, but it can become more relevant for the investigation of nonlinearities at low pump levels, when it is important to curtail nonlinear effects induced by the probe.

We conclude our discussion with a word of caution on the possibility of achieving quantum-enhanced measurement in a metrological sense. As a matter of fact, our investigation reveals that the modification of the mode prevents from casting the problem as a simple loss estimation problem~\cite{PhysRevLett.98.160401}, at odds with the static conditions~\cite{refId0}.





\section{Results}

The idea of quantum illumination is to exploit the correlations present in the emission of parametric processes in nonlinear materials. Two beams, termed signal and idler, are generated simultaneously, thus the presence of a photon on one mode heralds the presence of a photon on the other with high probability. In such a photon counting experiment, the intra-beam correlation is captured by means of the coefficient  
\begin{equation}
    g_{s-i}=C_{s-i}\frac{R}{S_iS_s}=\frac{C_{s-i}}{C_{\rm acc}},
\end{equation}
where $S_i$ ($S_s$) is the count rate of the idler (signal) arm, $C_{s-i}$ is the coincidence rate between the two arms, and $R$ is the repetition rate of the driving laser, in the pulsed regime. This coefficient also equals the ratio of the registered coincidence rate to the rate $C_{\rm acc}$ of the accidental events. Values of $g_{s-i}$ exceeding 100 are routinely obtained.

The signal arm is used to monitor an object. If $S_s$ is polluted by stray light events, one isolate their constribution to $S_s$ considering $S_s=S_t+S_n$, with $S_t$ the 'true' downconversion events, and $S_n$ the uncorrelated noise; the idler rate $S_i$ is assumed to be in a location with low noise. As a result, The observed coincidence rate is approximately
$C_{s-i}=g_{s-i}\frac{S_iS_t}{R}+\frac{S_iS_n}{R}$: its signal-to-noise ratio (SNR) is enhanced by a factor $g_{s-i}$, with respect to that of the single rate $S_s$. 

\begin{figure}[t!]
\centering
\includegraphics[width=\columnwidth]{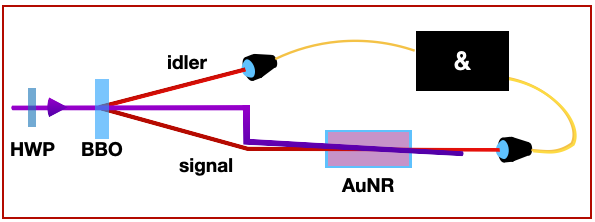}
\caption{Basic blocks of the pump and probe setup. A nonlinear $\beta$-barium borate (BBO) cristal is pumped by a blue beam in order to produce a pair of photons, labelled idler and signal. The idler is used as a trigger, while the signal monitors a suspension of gold nanorods. This is illuminated by the same pump beam generating the pair. Photons are detected by avalanche photodiodes, and their coincidences are registered by proper electronics.}
\label{fig:setup}
\end{figure}

Figure~\ref{fig:setup} sketches the main elements of our experimental setup for observing thermal lensing. The idler beam is sent directly to a single-photon detector, the signal beam reaches a suspension of AuNR subject to a pumping process. Coincidences betwee signal and idler photons are registered every $0.1$ s, with the pump activated by a manual shutter around $t=40$s in order to collect a reliable baseline signal.  

\begin{figure*}[h!]
\centering
\includegraphics[width=\textwidth]{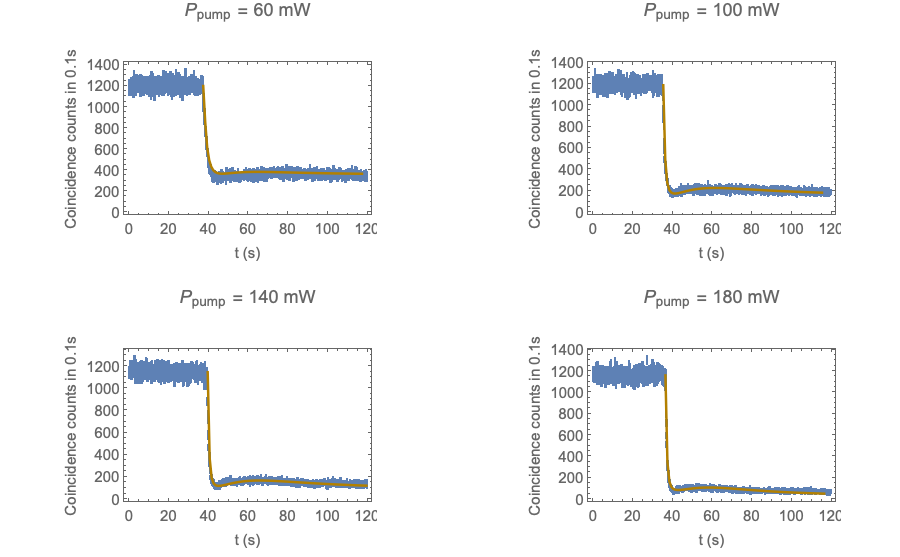}
\caption{Effect of pumping. Coincidence counts as a function of time at different pump levels, taken as the pump is injected on the beam. Error bars are calculated assuming Poissonian statistics. The solid line is a model based on Ref~\cite{Malacarne:11}}
\label{fig:pumpon}
\end{figure*}

The presence of the pump induces a change in the refractive index of the suspension, giving origin to a thermal lens effect that evolves over time. The registered coincidence counts are reported in Fig.~\ref{fig:pumpon} as a function of time at different pump power measured before the sample. We observe how the thermal effect produces a reduction of the signal. This is consistent with the negative Kerr effect reported in previous investigations, with the remarkable difference that both pump and probe beams are initially collimated, rather than focussed in the middle of the sample.    

The lens effect sets almost completely in a short time. Before reaching its steady state, the signal shows a small recoil and then a further slow decrease at longer times. For an explanation of such a behaviour, we can adopt the model in Ref.~\cite{Malacarne:11}: this describes the effect of the pump by means of a time-dependent spatial phase $\phi(r,t)$ ($r$ is the radial coordinate of the beam). Its expression consists of two terms
$\phi(r,t)=\phi_{\rm th}(r,t)+\phi_{\rm Soret}(r,t)$, where 
\begin{equation}
\begin{aligned}
    \phi_{\rm th}(r,t) = \frac{\theta_{\rm th}}{t_{\rm th}}\left[c_r \int_{t_0}^t \frac{1-\exp\left[-\frac{2m g}{1+2t'/t_{\rm th}}\right]}{1+2t'/t_{\rm th}}dt'\right.\\
    +\left.e^{-k t}(1-c_r) \int_{t_0}^t e^{k t'}\frac{1-\exp\left[-\frac{2m g}{1+2t'/t_{\rm th}}\right]}{1+2t'/t_{\rm th}}dt'\right],
    \label{eq:dasmodel}
\end{aligned}
\end{equation}
where we have introduced the rescaled variables $g=r^2/w_s^2$ and $m=w_s^2/w_p^2$, with $w_s$ ($w_p$) the beam radius of the signal (pump). A similar effect holds for the second term $\phi_{\rm Soret}(r,t)$, up to the replacements $\theta_{\rm th}\rightarrow \theta_{\rm S}$ and $t_{\rm th}\rightarrow t_{\rm S}$.

The phase $\phi_{\rm th}(r,t)$ describes the thermal contribution to the variation of refractive index induced by heating from the pump. The term $\phi_{\rm Soret}(r,t)$, instead, is linked to the Soret effect, determining a change in concentration, which takes place over a characteristic time $t_{\rm S}\gg t_{\rm th}$. In addition, there may occur a removal of population with an exponential decrease which is attributed to photochemical effects in aqueous solutions of molecules~\cite{Malacarne:11}, decreasing the concentration of the active molecules to an equilibrium value $c_r$. Its origin in the present case can not be attributed with the same clarity.  

From Fresnel diffraction theory~\cite{SHEN1992385}, it is possible to use the expression of $\phi(r,t)$ in order to predict the value of the intensity at the detection location~\cite{Malacarne:11}. We have fixed the measured value of $m$, and used $t_{\rm th}, t_{\rm S}, \theta_{\rm th}, \theta_{\rm S}$, and $k$ as free parameters in order to obtain the solid curves in Fig.~\ref{fig:pumpon}, as guides - a complete fitting procedure proved to be too demanding in terms of computing time. Remarkably, all three time scales $t_{\rm th}, t_{\rm S}, k$ were necessary to reproduce the temporal features of the data within this model. The resulting description allows us to identify the initial, faster decrease with the thermal effect, and the slow decrease at later time with the Soret effect.

Inspection of the data shows that the thermal response becomes faster at higher pump powers, while the overall effect is prone to saturation: notice the relatively smaller difference between the data at higher powers.

\begin{figure*}[h!]
\centering
\includegraphics[width=\textwidth]{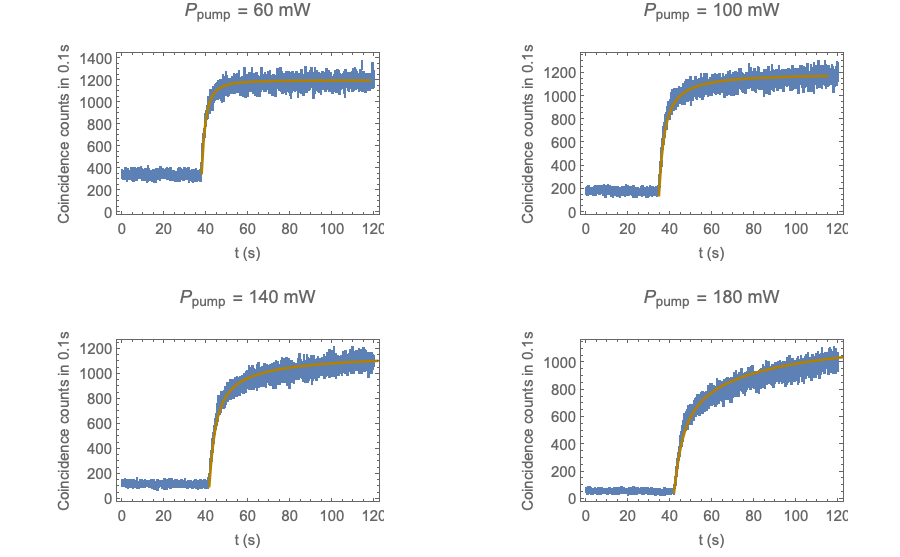}
\caption{Re-equilibration of the AuNR suspension. Coincidence counts as a function of time at different pump levels, taken as the pump is removed after $\sim$300 s illumination. Error bars are calculated assuming Poissonian statistics. The solid line is the result of a model described in Ref~\cite{Malacarne:11}}
\label{fig:pumpoff}
\end{figure*}

We can corroborate this description by looking at the reverse process in which the pump is removed. We proceeded similarly to the previous case, by blocking the manual shutter after about $40$ s, and observed the variation of the coincidence counts, as shown in Fig.~\ref{fig:pumpoff}. The solid curves are obtained with the same model described above, and the characteristic double rise time validates the presence of a Soret contribution, along with contribution from the thermal effect. The estimated response times in this case tend to be longer, revealing a somehow expected asymmetry between excitation and relaxation.

\begin{figure}[h!]
\centering
\includegraphics[width=\columnwidth]{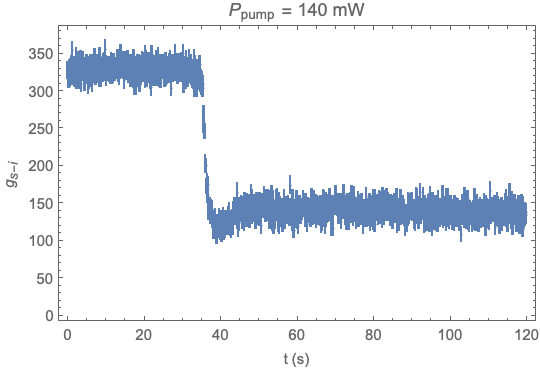}
\caption{An example of measured cross-correlation coefficient $g_{s-i}$ as a function of time. Its resilience to loss suggests that the observed change can not be attributed to a mere variation of the transmission.}
\label{fig:gtwo}
\end{figure}

The intensity correlation between signal and idler beams has been considered to deliver a metrological advantage in the estimation of transmittance~\cite{PhysRevLett.98.160401,PhysRevA.79.040305,Losero,refId0}. However, these realisations tipically consider a fixed geometry, ensuring that the value of correlation $g_{s_i}$ remains unaltered with respect to the initial, optimised level. This circumstance is key to performing a meaningful estimation. In our case, the modal structure of the signal beam is modified as a function of time. Further, the downconversion emission, observed in the single counts, is known to exhibit a more multimode profile than in the coincidences~\cite{PhysRevA.72.062301}. As a result, the correlation coefficient $g_{s-i}$ changes in time: we report an example in Fig.~\ref{fig:gtwo} demonstrating a variation by over a factor 2. 

\section{Discussion}
We have investigated thermal lensing in a suspension of AuNR by employing quantum light, as inspired by  quantum illumination. The observed profiles do ressemble what has been reported in the literature and point to the presence of multiple effects contributing to the nonlinearity~\cite{Malacarne:11}. The technique thus provide reliable information on the dynamics of the suspension on its way out of equilibrium or back into it. 

Perspectives of these studies may consider applications in confined environments, such as hollow fibres or microfluidics structures,  where nonlinearities can be achieved with lesser pump power due to the mode confinement: the employ of low-intensity probes can be more beneficial.

\section{Methods}

{\bf Synthesis of the AuNRs} AuNRs are produced by the two-step method described in Ref.~\cite{Amatori}. First, a seed solution is prepared, which is then mixed with silver nitrate, auric tetrachloric acid and ascorbic acid, thus yielding to the growth of nanorods. Purification is achieved by centrifugation at 1300 rpm for 10 minutes, repeated twice. Typical dimensions range are 10$\pm$2 nm and $40\pm$6 nm for the two characteristic dimensions of the rods. The acqueous solution employed in the measurements has a concentration of 1 mg/mL.

{\bf The quantum light source} Photon pairs are produced by parametric down conversion. The pump beam at 403.5 nm is obtained by a second harmonic generation in a 1-mm $\beta$-barium borate (BBO) nonlinear crystal, from a fundamental beam at 807 nm, with 160 fs duration and 80 MHz repetition rate. The total power of the blue pump can be adjusted by setting that of the fundamental. This part is not shown in Fig.~\ref{fig:setup}.

The pump beam reaches a second BBO crystal in which non-collinear, degenerate parametric down conversion occurs. The pump is focussed by a $f =152$ mm parabolic mirror, while the photons are collimated by a similar mirror (not shown in Fig.~\ref{fig:setup}). The bandwidth of the photons is limited to 7.5 nm (FWHM) around 807 nm by interference filters. The coincidence level at the baseline is kept at 1200 events in 0.1 s in order to limit multipair emission. 
Spatial collection employs single-mode fibre with 10x microscope objectives. Detection efficiencies are of the order of 6\%, estimated by the Klyshko method~\cite{klyshko80sjqe}. The pump is also collimated at the output by means of a lens with focal length $f=150$ mm.

The measurement of the ratio $m$ was carried out by a direct measurement on a camera for the pump beam ($w_p=0.57$ mm). For the signal beam, we have measured the detected mode by shining an 810 nm laser backwards in the detection fibre ($w_s=0.61$ mm).

The efficiency of the down conversion process depends on the polarisation of the pump: this is maximal when the polarisation is aligned along the extraordinary direction, while the process is inhibited for the ordinary polarisation. The control by a half-waveplate (HWP) allows to keep the same coincidence level for all powers (Fig.~\ref{fig:setup}). Since the AuNR solution is at a relatively low concentration (1mg/mL), we expect no effects related to the polarisation.

The AuNR suspension is contained in a 20 mm glass cuvette, set in the direction of the signal by means of a reference laser following the same path. This is illuminated by the pump beam oriented at an angle of $3.2^\circ$ with respect to the signal probe ($\cos(3.2^\circ)>0.998$).

\section{Author information}
{\bf Corresponding Author}\\
*Email: ilaria.gianani@uniroma3.it\\
{\bf Notes}\\
The authors declare no competing financial interest.

\begin{acknowledgement}
The authors thank A. Datta for discussion.
This work was supported by the European Commission (FET-OPEN-RIA STORMYTUNE, Grant Agreement No. 899587).


\end{acknowledgement}



\bibliography{mainbib}

\end{document}